\documentclass[amsmath, aps]{revtex4}
\usepackage{graphicx}
\usepackage{times}
\usepackage{epsfig}
\usepackage{amssymb}
\usepackage{amsmath}

\begin{document}
\title{Answer on the comment by S. G. Karshenboim, V. G. Ivanov and J. Chluba}

\author{L. Labzowsky$^{1,2}$ and D. Solovyev$^1$}

\affiliation{ 
$^1$ V. A. Fock Institute of Physics, St. Petersburg
State University, Petrodvorets, Oulianovskaya 1, 198504,
St. Petersburg, Russia
\\
$^2$  Petersburg Nuclear Physics Institute, 188300, Gatchina, St.
Petersburg, Russia}
\maketitle

We studied carefully the comment by S. G. Karshenboim, V. G. Ivanov and J. Chluba \cite{KIC} to our paper "Two-photon approximation in the theory of the electron recombination in hydrogen" \cite{SL} and found this comment erroneous.

1. The errors begin with the second formula in \cite{KIC} which reads
\begin{eqnarray}
\label{1}
\Gamma_A=W_{A-1s}^{\rm total},
\end{eqnarray}
where $\Gamma_A$ is the total radiative width (i.e. total decay probability of state A), $W_{A-1s}^{\rm total}$ is the sum of probabilities for the different decay channels $A-1s$. Strictly speaking, no such formula does exist. Eq. (\ref{1}) has no sense until the type of the experiment is specified. The standard definition of the one-photon width of the level $\Gamma_A^{1\gamma}$ is 
\begin{eqnarray}
\label{2}
\Gamma_A^{1\gamma}=\sum\limits_{A'\,\,(E_{A'}<E_A)}W_{A-A'}^{1\gamma},
\end{eqnarray}
where $W_{A-A'}^{1\gamma}$ are the one-photon transition probabilities to the lower levels. Eq. (\ref{2}) can be strictly derived from QED assuming that $\Gamma_A^{1\gamma}$ can be defined as an imaginary part of the lowest order electron self-energy correction $\Delta E_A^{SE}$. The idea belongs to Barbieri and Sucher \cite{Bar} and the  derivation one can find, for example, in \cite{LSP} and in more detail in \cite{LKD}.

In the laboratory experiments usually the one-photon transitions $W_{A-A'}^{1\gamma}$ are measured and the level width is defined according to Eq. (\ref{2}). The level widths defined in this way are commonly given in the textbooks.

Of course there are exceptional cases, like the level 2s in hydrogen when the width is defined by the two-photon transition.

The different situation arises when we want to define the transition rate from the level A to ground state via the cascade transition. This is directly connected with the recombination process and this is the question which we wanted to answer in our paper \cite{SL}. In this case the decay rate  (if there is only one decay channel) may not coincide with the level width $\Gamma_{A}^{1\gamma}$ measured via detecting photons with definite frequency as in the experiments of the first type, described above.

The difference between the two types of the experiments can be well understood on the simplest example: decay of the 3s level. To the ground state this level (neglecting all magnetic and electric multipole transitions) can decay via the unique cascade $3s\rightarrow 2p\rightarrow 1s$. In the laboratory experiment of the first type one can detect the photons with frequency $\omega=E_{3s}-E_{2p}$ and measure the transition rate $W_{3s-2p}^{1\gamma}$ which equals to $W_{3s-2p}^{1\gamma}=6.317\cdot 10^6$ $s^{-1}$. Then the width $\Gamma_{3s}$ should be defined as
\begin{eqnarray}
\label{3}
\Gamma_{3s}^{1\gamma}=W_{3s-2p}^{1\gamma}=6.317\cdot 10^6\,\,s^{-1}.
\end{eqnarray}
This is the value which is presented in textbooks.

If we consider the experiment of the second type (recombination) we have to evaluate the differential two-photon transition rate $dW_{3s-1s}^{2\gamma}(\omega)$ and then obtain the total transition rate as
\begin{eqnarray}
\label{4}
W_{3s-1s}^{2\gamma}=\frac{1}{2}\int\limits_0^{E_{3s}-E_{1s}}dW_{3s-1s}^{2\gamma}(\omega).
\end{eqnarray}
This evaluation was performed in \cite{LSP} with the result 
\begin{eqnarray}
\label{5}
W_{3s-1s}^{2\gamma}=6.285\cdot 10^6\,\, s^{-1}
\end{eqnarray}
The values defined by Eq. (\ref{3}) and Eq. (\ref{5}) do not coincide. Thus the equality (\ref{1}) does not hold in this case. The transition 3s-1s via the cascade is a bit slowlier than the one-photon transition 3s-2p. This can happen due to the destructive interference with the "pure" two-photon contribution to $dW_{3s-1s}^{2\gamma}(\omega)$. In case of $3d\rightarrow 1s$ two-photon emission the values of $\Gamma_{3d}^{1\gamma}$ and $W_{3d}^{2\gamma}$ differ more seriously:
\begin{eqnarray}
\label{5.1}
\Gamma_{3d}^{1\gamma}=W_{3d-2p}^{1\gamma}=6.469\cdot 10^7\,\, s^{-1},
\end{eqnarray}
\begin{eqnarray}
\label{5.2}W_{3d-1s}^{2\gamma}=5.814\cdot 10^7\,\, s^{-1}.
\end{eqnarray}

Now let us look what happens in case of the 3p level decay, expressed by the formula (33) in \cite{SL}. This formula was strongly criticized in \cite{KIC}. In this case there is direct one-photon decay to the ground state, so in the experiment of the first type the photons with the frequencies $\omega_1=E_{3p}-E_{1s}$ or $\omega_2=E_{3p}-E_{2s}$ can be detected and the total width of the 3p level according to Eq. (\ref{2}) is
\begin{eqnarray}
\label{5a}
\Gamma_{3p}^{1\gamma}=W_{3p-1s}^{1\gamma}+W_{3p-2s}^{1\gamma}.
\end{eqnarray}
The decay rate $W_{3p-1s}^{\rm total}$ in case of 3p level is defined by Eq. (33) in \cite{SL}. This equation reads
\begin{eqnarray}
\label{5b}
W^{\rm total}_{3p-1s}=W_{3p-1s}^{1\gamma}+\frac{3}{4}W_{3p-2p}^{2\gamma}+\frac{3}{4}\frac{W_{3p-2s}^{1\gamma}}{\Gamma_{3p}}W_{2s-1s}^{2\gamma}.
\end{eqnarray}

The first term in Eq. (\ref{5b}) is, of course, dominant and by the order of magnitude is close to (\ref{5a}). Two other terms represent the contributions of two cascades 3p-2p-1s and 3p-2s-1s, i.e. $3\gamma$ transitions. Both these contributions are quite small, since the two-photon links of the cascade transitions from 3p to 1s are very slow. The transition rate via cascade is defined by the slowest link of the cascade, in particular by the two-photon links 3p-2p or 2s-1s. This result automatically arises in our line-profile QED approach and is reflected in Eq. (\ref{5b}). This approach appears to be useful for the analysis of the different types of the two-photon transitions ("two-photon approximation").

In \cite{KIC} the authors discuss mainly the third term in the right-hand side of Eq. (\ref{5b}). In their Table I they suggest to replace it by the $W_{3p-2s}^{(1\gamma)}$ and to define this value as the partial width of the $3\gamma$ decay. It is true that if the photon with frequency $\omega=E_{3p}-E_{2s}$ is detected, the decay to the ground state is necessarily $3\gamma$ decay. However, in the recombination experiment we are interested in the transition rate $3p-1s$ via 3-photon decay. This transition rate is given by the second and third terms in the right-hand side of Eq. (\ref{5b}).

The same picture (in a more complicated form) occurs for the cascade decays of the levels 4s,4d discussed in \cite{SL}. The difference between $\Gamma^{1\gamma}_A$ and $W^{\rm total}_{A-1s}$ was not recognized by the authors of \cite{KIC}. No wonder that they found the difference between these quantities in their Tables I and II in \cite{KIC}.

In our paper \cite{SL} another question which is related to astrophysics was also discussed. This is a question of the "radiation escape" from the matter. We considered this problem in a very simplified formulation: every two-photon transition contributes to this escape. So let's look at Eq. (\ref{5b}) from this point of view. Now we want to know how probable is the two-photon emission in this process. Analysis of Eq. (\ref{5b}) helps us to answer this question too. The third term in the right-hand side of Eq. (\ref{5b}) means that with the probability $W_{3p-2s}^{(1\gamma)}/\Gamma_{3p}$ (this is branching ratio for the transition 3p-2s) the electron will arrive at the level 2s, where it has no choice but to go down to 1s state, emitting two photons. This contribution to the radiation escape is, of course, taken into account in the astrophysical balance equations. The contribution of the second term in the right-hand side of Eq. (\ref{5b}), i.e. $W_{3p-2p}^{(2\gamma)}$ was not yet discussed in the context of the radiation escape. However, this contribution should be quite small due to the small branching ratio $W_{3p-2p}^{(2\gamma)}/\Gamma_{3p}$. The same concerns the two-photon decay channels in Eq. (47) in \cite{SL}: their contributions are also negligible. Still it does not mean that such contributions are always negligible: the particular examples will be given in a subsequent paper.

2. Another error concerns the method of regularization of the resonance denominators. There is a well-defined unambigous procedure first formulated in QED by F. Low \cite{Low} and later applied to the cascade processes in \cite{LabShon}. According to this approach, an infinite number of the lowest-order electron self-energy corrections should be inserted in the resonance approximation itno Feynman graph, corresponding to the two-photon decay (see Fig. 1). These insertions can be converted to the geometric progression. Then the graphs Fig. 1a), 1b) produce the energy denominators, containing $\Gamma^{1\gamma}_{3s}+\Gamma^{1\gamma}_{n}$, where n denotes the resonance intermediate state. Summation of singular terms, corresponding to the insertions in Fig. 1c) should be done in a different way. The singularities in Fig. 1c) are not connected with the resonance and can be avoided only by introducing the Gell-Mann and Low \cite{Gell} adiabatic parameter $\alpha$ and finally evaluating the limit $\alpha \rightarrow 0$. It can be shown \cite{Andr} that the summation of the singular graphs converts to the factor $exp(-i\Delta E_{1s}^{SE}/\alpha)$ in the amplitude. For the probability of transition to the ground state this gives
\begin{eqnarray}
\label{6}
\lim\limits_{\alpha\rightarrow 0}\left | e^{-i\Delta E_{1s}^{SE}/\alpha}\right|^2=1.
\end{eqnarray}
Moreover, with this procedure the Lamb shift for the ground state also arrives in the energy denominators \cite{Andr}. 

The same approach can be unambigously applied to the $3\gamma$, $4\gamma$ etc transitions. However, this clearly formulated procedure does not satisfy the authors of \cite{KIC}. They do not refer neither to \cite{Low} not to \cite{LabShon}. Instead they refer to the paper \cite{Karsh} by S. G. Karshenboim and V. G. Ivanov, where the problem of regularization of the resonant terms is considered at the quantum-mechanical (QM), i.e. phenomenological level. This eventually leads to the crucial errors. For example, the authors of \cite{KIC} insist that the initial width should not enter the energy denominators. In the QED picture it is evident that it should since otherwise we have to omit for some unknown reasons the contribution of the graphs Fig. 1a).

In the paper \cite{Karsh} the eigenmode decomposition of the Green function was used for the description of the sum over intermediate states. This decomposition containes the denominators with the phenomenological width as a parameter. The authors of \cite{Karsh} attribute this parameter fully to the intermediate state $\Gamma_n^{1\gamma}$, though, according to QED, in case of cascade transition it should be attributed to the sum $\Gamma^{1\gamma}_{i}+\Gamma^{1\gamma}_{n}$, where i is the initial state. The same is true for the every link of the cascade as it was explicitly proved with QED approach in \cite{LabShon}.

This error demonstrates the danger of using the phenomenological QM theory instead of QED for the description of the complicated processes in atoms. It is especially regrettable since it can deteriorate in future the accurate astrophysical codes.

3. We turn now to another conceptual error contained in \cite{KIC}. This error concerns considerations about the validity of the theory developed in our paper \cite{SL}, i.e. the theory of the cascade transitions in the resonance approximation. The authors of \cite{KIC} draw the conclusion that these validity conditions are
\begin{eqnarray}
\label{7}
\frac{\Gamma_{\rm initial}}{\Gamma_{\rm intermediate}}\ll 1 \,\,\,{\rm and}\,\,\, \frac{\Gamma_{\rm final}}{\Gamma_{\rm intermediate}}\ll 1,
\end{eqnarray}
where the widths of the initial, final and intermediate states are involved. We should stress that the conditions Eq. (\ref{7}) contradict to the basic principles of QED and, in general, of Quantum Field Theory (QFT). The main tool of the QFT is the S-matrix which is constructed in such a way that one has to fix only initial and final states (stable or metastable) and then perform the uniquely defined evaluations that result in the transition probability (cross-section). No additional information about the intermediate states for any cascade (or not cascade) process is required. All this information is contained in the S-matrix. The widths of the intermediate states in the cascade process will arise automatically via the summation of the self-energy insertions, described above (see Fig. 1). No conditions of the type Eq. (\ref{7}) are necessary.

All the S-matrix elements are finite provided that the standard renormalization procedure is performed. The divergent integrals in the cascade transition matrix elements arise only due to the improper use of the QED perturbation theory and can be eliminated by the resummation of the self-energy graphs. In no way the properties of the intermediate states can influence the validaty of the general QED approach to the evaluation of the S-matrix elements for any transitions. Of course, in astrophysics the situation is much more complicated than for the free atoms, considered in \cite{SL}, but the basic principles should not be violated in any case.

The authors of \cite{KIC} also raise the question about the dependence of the initial state on creation process and connect this question with the validity of the cascade (resonance) approximation. In principle, the QED treatment of any process is absolutely well-defined only when this process starts with the stable (ground) state and ends up with this state too. Then no question of preparation can arise. Still it is possible also to consider the metastable states as initial ones. Note that in this context all the atomic states can be considered as metastable since the energy level widths are always smaller than the intervals between the levels. Of course, the formulation of the problem of the transition probability from a certain initial excited state is always an idealization. This formulation is possible provided that in the process of creation of the initial state the so-called nonresonant (NR) corrections can be treated as negligible. The nonresonant corrections were first introduced by F. Low \cite{Low} and during the last decade were studied for different atomic processes \cite{LSPS}-\cite{LSSCK}. The NR corrections can depend on the creation story. However, the general conclusion after the series of works \cite{LSPS}-\cite{LSSCK} was that they are extremely small for all types of the excitation (i.e. creation) processes. Hence, we can safely assume that the creation story of the initial state in no way affects the cascade transition evaluations.

Concluding, we can state that, contrary to the statement in \cite{KIC}, all the formulas and the results in our paper \cite{SL} are quite correct. As it was formulated in \cite{SL} "The main goal of our paper is the formulation of the two-photon approximation which allows for the rigorous incorporation of all types of two-photon processes. This may be important for more accurate astrophysical investigations of the cosmic radiation background". Some applications of the results of paper \cite{SL}, more close to astrophysics will be presented in a subsequent paper.

\begin{figure}[htp]
\includegraphics[scale=0.6]{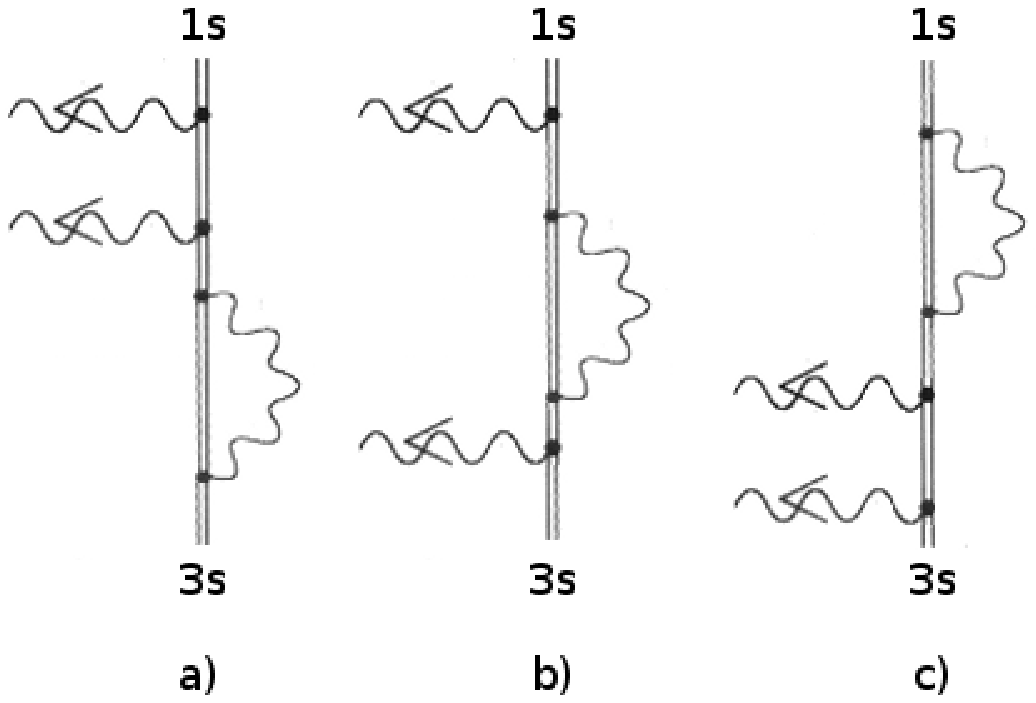}
    \\
{\citation
\\ Fig. 1. Electron self-energy insertions into the initial, intermediate and final states for the two-photon 3s-1s transition. The double solid line represents the electron in the field of the nucleus, the wavy lines represent the emitted ot virtual photons. The electron self-energy loop represent the infinite chain of SE insertions in the resonance approximation. The insertions in Fig. 1a) lead to the arrival of the initial level width $\Gamma_{3s}$ in the resonance energy denominators, the insertions in Fig. 1b) correspond to the intermediate level width and the insertions of the type 1c) produce the Lamb shift (pure real) for the ground state 1s.}
\end{figure}
\end{document}